# Effects of Inflow Turbulence on Structural Deformation of Wind Turbine Blades

Linyue Gao [1], Shu Yang [1], Aliza Abraham [1,2], Jiarong Hong [1,2,*]

[1] St. Anthony Falls Laboratory, University of Minnesota, Minneapolis, MN 55414, USA

[2] Department of Mechanical Engineering, University of Minnesota, Minneapolis, MN 55455, USA

[*] Corresponding author: jhong@umn.edu (J. Hong).

## Abstract

A better understanding of the intense interaction between the turbulent inflow and rotating wind turbine components is critical to accelerating the path towards larger wind turbines. The present investigation provides the first field characterization of the influence of turbulent inflow on the blade structural response of a utility-scale wind turbine (2.5MW), using the unique facility available at the Eolos Wind Energy Research Station of the University of Minnesota. A representative one-hour dataset under a stable atmosphere is selected for the characterization, including the inflow turbulent data measured from the meteorological tower, high-resolution blade strain measurement at different circumferential and radiation positions along the blade, and the wind turbine operational conditions. The results indicate that the turbulent inflow modulates the turbine blade structural response in three representative frequency ranges: a lower frequency range (corresponding to modulations due to large eddies in the atmosphere), a higher frequency range (corresponding to flow structures in scales smaller than the rotor diameter), and an intermediate-range in between. The blade structure responds strongly to the turbulent inflow in the lower and intermediate ranges, while it is primarily dominated by the rotation effect and other high-frequency characteristics of wind turbines in the higher frequency range. Moreover, the blade structural behaviors at different azimuth angles, circumferential and radial locations along the blade are also compared, suggesting the comparatively high possibility of structural failure at certain positions. Further, the present study also uncovers the linkage between the turbulent inflow and blade structural response using temporal correlation. The derived findings provide insights into the development of advanced control strategies or blade design to mitigate the structural impact and increase blade longevity for the safer and more efficient operation of large-scale wind turbines.

# 1. Introduction

Wind power is considered one of the most promising alternative source of energy worldwide and has achieved competitive or even lower prices than traditional fossil fuels in the United States [1]. As wind turbines continue to grow in size with longer blades for more efficient power generation, the interactions between turbulent inflow and wind turbines become more intense and, consequently, have stronger effects on wind turbine performance [2]. The turbulent inflow can result in dramatic fluctuations in wind turbine power output and thereby may cause unexpected frequency deviations to the power utilities [3,4]. More importantly, wind turbine structures, particularly the rotating blades, are susceptible to increased loads due to higher levels of turbulence [5]. In recent years, the varying turbine loading due to atmospheric turbulence has been widely regarded as an important factor that needs to be fully considered to achieve the 20-year design life criterion of wind turbines, especially utility-scale wind turbines [6]. Currently, most studies of the effect of turbulence on wind turbines focus on power fluctuations [2,4,7–9], while the ones on flow-blade interaction focus on blade health monitoring based on long-term fatigue loading [10–13]. However, it is also critical to investigate the effect of turbulence on blade structural response in the short term to provide insights into developing advanced control strategies for load/deformation mitigation for safer and more efficient wind turbine operation [12,14].

In general, very few studies have provided a detailed examination of the effect of turbulence on blade structural response due to the limitations in simulation methods and field/experimental facilities [2]. Only in recent years, several numerical investigations [10,11,15–20] have been conducted to look into this subject for utility-scale wind turbines. Specifically, Lee et al. [12] investigated blade loading of the NREL 5MW wind turbine under turbulent inflow simulated using large eddy simulation (LES) and blade deformation captured using an aeroelastic code, FAST. They found that the blade root moment showed a strong energy peak at the turbine rotor frequency (i.e., $f_T$ = 1P), in contrast to the spectra of the tower yaw moment which peaked at the blade passing frequency (i.e., $f_{BPF}$ = 3P). Following this work, Churchfield et al. [13] from the same group showed that the flapwise blade root bending moment and low-speed shaft torque are well-correlated with the incoming wind with a time offset matching the advection time of the flow from the inflow sampling location to the turbine rotor plane. The same turbine was simulated by Vijayakumar et al. [21] via a hybrid URANS-LES scheme and an actuator line model (ALM). Temporal variation in blade loading was found to be associated with three distinct time scales, corresponding to the advection of atmospheric flow structures through the rotor, rotor rotation and turbulence-induced forcing as the blades traverse internal atmospheric flow structures. Sabale et al. [18] used an advanced aeroelastic code, WindGRAR, and LES to simulate the same wind turbine, and provided more information on the effects of transverse turbulence, wind shear and flow unsteadiness on the aeroelastic response of turbine blades. Besides the NREL 5MW reference wind turbine, a 33-kW 2-bladed wind turbine was simulated by Li et al. [22] by using LES and FAST for blade deformation. Their results indicate that the flapwise blade root bending moment has a clear scale-to-scale response to the turbulent structure of the atmosphere based on wavelet analysis. The aforementioned simulations provide valuable information on blade structural response under turbulent inflow, particularly at the blade root region. However, these simulations usually involve a number of simplifications on the mechanical/aerodynamic characteristics of the blade (e.g., 3D flow effects) and neglect the influence of the tower and nacelle on blade motion. They also lack validation from field experiments due to the unavailability of well-equipped utility-scale test facilities and challenges of high-resolution characterization of turbulent inflow and blade structural response. These drawbacks limit their applicability to utility-scale turbines.

To date, Nandi et al. [8] in 2017 is the only field measurement that directly investigates the linkage between turbulent flow and blade response. In their study, the near-blade velocities were measured using pitot probes at blade leading-edge and trailing-edge positions of a GE 1.5 MW wind turbine under natural environments. The same three response time scales as those reported in Vijayakumar et al. [21] were experimentally identified. However, no direct measurements of blade loading and strain were obtained to link such turbulent inflow signatures with blade structural response. In addition, no systematic investigation has been conducted on blade structural response at different blade sections and azimuthal angle positions.

Consequently, the current study provides a systematic investigation of the turbulence-blade interaction of a utility-scale wind turbine, including the structural response of the turbine blades at different blade sections and azimuthal angle positions. Our investigation leverages the data collected at the Eolos Wind Energy Research Station at the University of Minnesota. This facility consists of a well-instrumented 2.5 MW wind turbine and a meteorological tower, allowing simultaneous characterization of turbine operational conditions, blade structural response, and turbulent inflow conditions. The detailed methodology of our investigation is described in Section 2, followed by the data analysis in Section 3 and a conclusion and discussion of the main findings in Section 4.

## 2. Methods

### 2.1 Experimental facilities

Our investigation utilizes the field data collected at the University of Minnesota Eolos Wind Energy Research Station (referred to as the Eolos station hereafter) in Rosemount, MN. The station consists of a 2.5 MW upwind, 3-bladed, horizontal-axis wind turbine (Clipper Liberty C96, referred to as the Eolos turbine hereafter) and a 130 m meteorological tower (referred to as the met tower hereafter) [23], as shown in Fig. 1(a). The met tower, located 170 m south (i.e., along the prevailing wind direction) of the Eolos turbine, is equipped with velocity (sonic, and cup & vane anemometers), temperature, and relative humidity (RH) sensors to characterize the incoming flow. Four high-resolution sonic anemometers (Campbell Scientific, CSAT3) with a sampling rate of 20 Hz are mounted at four representative elevations, i.e., rotor top tip (129 m), hub height (80 m), rotor bottom tip (30 m), and standard 10 m. Six cup & vane anemometers (Met One, 014-A) are installed three meters below each sonic anemometer, i.e., at 126 m, 77 m, 27 m, and 7 m, and at two other elevations, i.e., 102 m and 52 m, corresponding to the midspans of lower and upper blades, respectively.

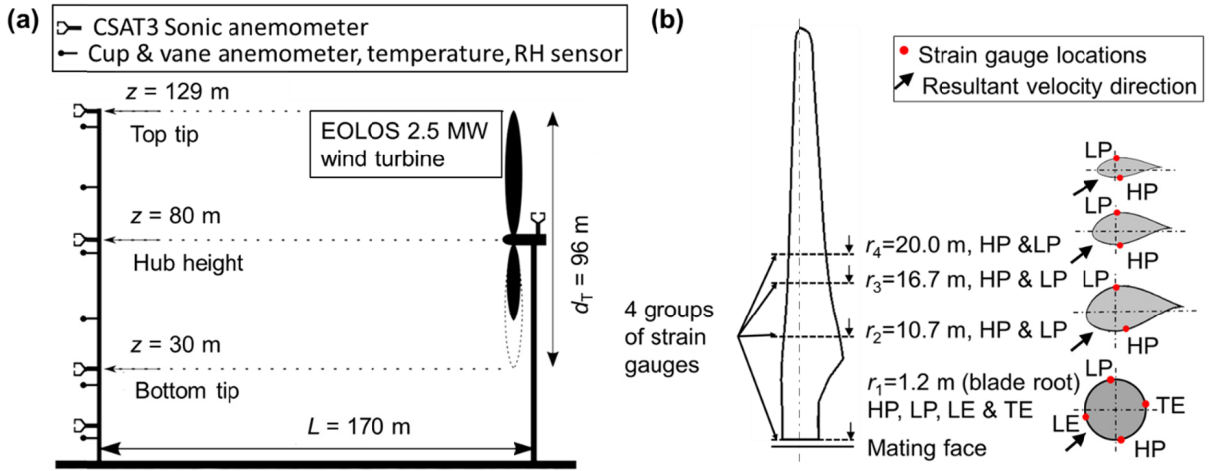

Figure 1: (a) Schematic of the 2.5 MW wind turbine and the meteorological tower at Eolos station. Adapted from Dasari et al. [23]. (b) Schematic of configurations of strain gauges installed on each Eolos turbine blade.

The Eolos turbine is variable-speed, variable-pitch regulated, with a rotor diameter of $d_T$ = 96 m, and a hub height of $z_{hub}$ = 80 m. The cut-in, rated and cut-out wind speeds are ~4 m/s, 11 m/s and 25 m/s, respectively. In addition to the standard Supervisory Control and Data Acquisition (SCADA) system used to record the turbine operating conditions at a sampling rate of 1 Hz, the turbine is instrumented with a high-resolution foundation sensor system and a blade sensor system to quantify the structural response at a sampling rate of 20 Hz. The foundation sensor system consists of 20 single-axis strain gauges (Vishay MicroMeasurements, CEA-06-125UN-350/P2), evenly distributed around the tower base. Detailed information on the foundation sensor system was reported by Chamorro et al [24]. Moreover, the turbine is instrumented with a blade sensor system consisting of 10 fiberoptic strain gauges (Micron Optics, os3200) installed at different radial positions along each blade (i.e., blade root (RT), 25% $R$, 37.5% $R$ and 45% $R$, where $R$ represents the radius of the turbine rotor, 48 m) for characterizing the blade deformation. The sensitivity of the strain gauges is ~1.2 pm/μm. As shown in Fig. 1(b), four strain gauges are installed over the high-pressure (HP) side surface, low-pressure (LP) side surface, leading-edge (LE) and trailing-edge (TE) regions of the cross-section at the blade root ($r_1$ = 1.2 m from the mating face). Six strain gauges are mounted over the HP and LP surfaces of the cross-sections at other outer radial positions, i.e., at 25% ($r_2$ = 10.7 m), 37.5% ($r_3$ = 16.7 m), and 45% ($r_4$ = 20.0 m) blade length away from the hub center, respectively. Temperatures measured at the same sampling rate for each strain gauge are used to compensate for temperature effects on strain measurements. The blade sensor system uses National Instruments Labview 2011 for data acquisition and Labview VI's (Virtual Instruments) designed for communication with the interrogator (Micron Optics, sm130) for data collection from strain gauges. The data from all the sensor systems at the Eolos station are recorded continuously since its establishment in 2011, constituting a 9-year database of turbine operation, turbine structural response, turbulent inflow, and other local meteorological parameters. Such a dataset has been used for a number of past studies on turbulence-turbine interaction [3,24] and turbine wake behavior [25,26].

**2.2 Dataset selection**

We examined the 9-year Eolos database to identify an appropriate dataset for investigating the blade structural response under turbulent inflow by applying a series of restrictions on the inflow condition, turbine operational condition, and data quality. In order to ensure the met tower can accurately probe the turbulent inflow, wind direction is set to be centered around the south, i.e., 160 ° < $\alpha$ < 200 °, where direct north is 0 °. Additionally, the yaw error of the Eolos turbine is limited to less than 20º. A series of restrictions on wind turbine operational conditions are applied to mitigate the effects of pitch control on blade structural response for the simplicity of the present study. The Eolos turbine is set to operate in the most efficient operational condition with the maximum power coefficient ($C_{p,\max}$ = 0.472) and small pitch angle (θ) (i.e., θ < 4º, and θ = 1 ° in the ideal condition). The corresponding hub-height wind speed ($U_{\text{hub}}$) ranges from 6.9 m/s to 9.2 m/s, while rotor speed ($\omega_T$) ranges from 11.69 RPM to 15.5 RPM. The datasets from the met tower, Eolos SCADA system, foundation sensor system, and blade sensor system for the selected intervals are set to have good data availability (> 98%) to ensure sufficient data for further analysis.

Based on the aforementioned restrictions, we filtered the data obtained from the Eolos database using a sliding window approach with a window size of one hour and an advancing increment of 10 minutes. It should be noted that the 20 Hz observations are reduced to 1 Hz by averaging consecutive sets of 20 data points during the filtering process. The atmospheric stability for each dataset is calculated based on Bulk Richardson number ($Ri_b$) and Monin-Obukhov length ($L_{\text{MO}}$). Only datasets with a stable atmosphere are selected for the present analysis because they can be used as the baseline case for turbulent inflow studies without convection effects. The qualifying datasets are ranked based on the standard deviation (std) of the wind direction, and the one with the smallest standard deviation (i.e., std($\alpha_{\text{hub}}$) = 3.7°) is selected for the following analysis.

This one-hour dataset is in the period ranging from 07:00:00.000 to 07:59:59.950 UTC time on 09/19/2013 (i.e., from 02:00:00.000 to 02:59:59.950 on 09/19/2013 local time). The inflow wind speed and wind direction measured by the sonic anemometer at hub height are quite stable during the selected period, as shown in Fig. 2(a) and (b). The mean wind direction at hub height ($\alpha_{\text{hub}}$) is 178 ° (i.e., almost southerly) with a small deviation of 3.7 ° during the selected period. The mean wind speed at hub height ($U_{\text{hub}}$) is 7.8 m/s. The mean horizontal wind speed ($U_h$) and wind direction at other elevations ranging from the standard 10 m to the top tip of the wind turbine rotor (i.e., 129 m) during the selected period are shown in Figure 2(c). The variation in wind direction across the rotor area (from 30 m to 129 m) is found to be quite small, i.e., Δα = 4.8°, indicating the effect of wind veer can be neglected. More detailed characteristics of the turbulent inflow measured at the met tower location during the selected period can be seen in Fig. 3. As the elevation increases, the mean temperature increases while the RH decreases, suggesting a stable atmospheric boundary layer, confirmed by $Ri_b$ = 0.026 and $L_{\text{MO}}$ = 81.60 m [27]. The data measured by the cup & vane and sonic anemometers have a good agreement in wind speed and turbulent intensity, as shown in Fig. 3(c) and (d). As the elevation increases, wind speed correspondingly increases while turbulence intensity decreases, indicating a strong wind shear condition. The mean turbulence intensity (TI) ranges from 13% (bottom tip) to 5% (top tip) across the turbine rotor. The mean TI and the dimensionless turbulent shear stress ($u'w'/U_{hub}^2$, where $u'$ and $w'$ are the two velocity components along the north-south and west-east directions, respectively) at hub height are approximately 7% and $2\times10^{-3}$, respectively.

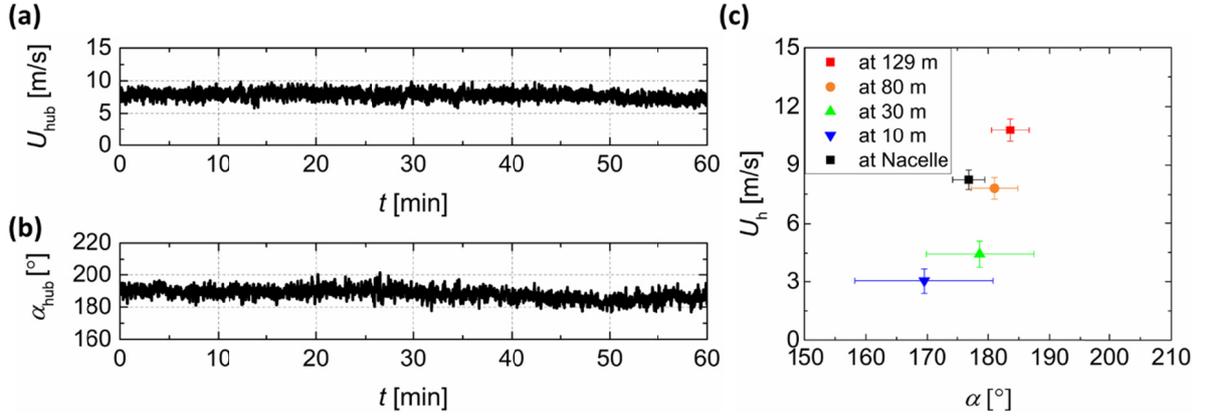

Figure 2: Inflow conditions measured by the sonic anemometers on the met tower during the selected period. (a) Time series of hub-height wind speed ($U_{hub}$); (b) time series of hub-height wind direction ($\alpha_{hub}$); and (c) mean horizontal velocity as a function of mean wind direction at various elevations.

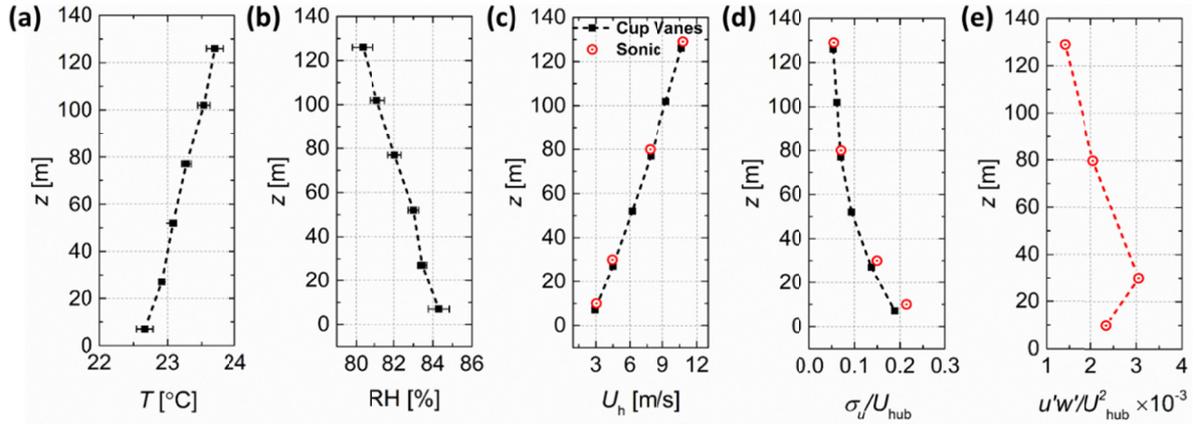

Figure 3: Characteristics of the turbulent inflow conditions based on met tower measurements during the selected period. (a) Mean temperature; (b) relative humidity; (c) mean velocity; (d) turbulence intensity; and (e) dimensionless turbulent shear stress.

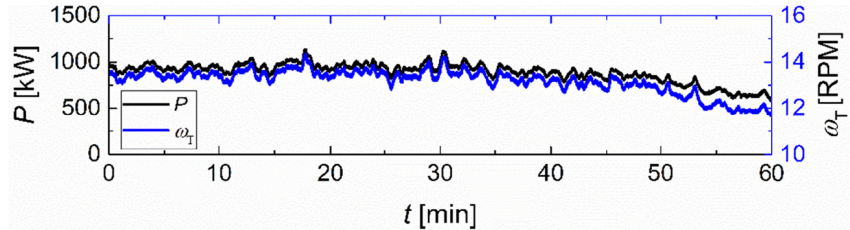

Figure 4: Time series of power ($P$) and rotor speed ($\omega_T$) of Eolos turbine during the selected period.

In addition, the mean turbine collective blade pitch angle and rotor speed ($\omega_T$) are 1° and 13.2 RPM, respectively, during the selected period, as shown in Fig. 4, indicating that the wind turbine operation meets the selection criteria. Based on the mean rotor speed, the turbine rotation frequency ($f_T$, i.e., 1P) and the blade passing frequency ($f_{BPF}$, i.e., 3P) are 0.22 Hz and 0.66 Hz, respectively, which will be used for the further analysis. Figure 4 also shows the time series of wind turbine power output ($P$). The power output fluctuates following a similar trend to the rotor speed in the selected period, as expected. The mean power is 892 kW during the selected period with a normalized variation of $\sigma_P/\bar{P} = 0.117$, where $\sigma_P$ is the root mean square (RMS) of the power fluctuations.

## 3. Results

### 3.1 Turbulent inflow effect on the structural response at the blade root

The effect of turbulent inflow on blade structural response is first investigated using the edgewise and flapwise blade bending moments measured at the blade root. Based on the difference between the blade strain measurements at LE and TE, and the HP and LP regions of the cross-section at blade root, the edgewise ($M_{Edge}$) and flapwise ($M_{Flap}$) blade bending moments are calculated, respectively. In the original data, the mean flapwise blade root bending moment is around 10 times larger than that in edgewise direction, indicating the flapwise moment dominates the structural response of the blade. The edgewise and flapwise bending moments are normalized by their maximum values (i.e., max($M_{Edge}$) and max($M_{Flap}$), respectively) during the selected period, as shown in Figure 5. The normalized edgewise bending moment varies from -1.0 to 1.0 (Fig. 5a). The positive and negative values indicate the moments along and against the rotation direction of the blade, respectively. During the selected period, the normalized edgewise blade root bending moment is relatively stable. As shown in the zoomed-in plot, a clear periodicity is observed, which corresponds to the rotation frequency ($f_T = 0.22$ Hz, i.e., 1P). The normalized flapwise blade root bending moment is found to be in the range of 0.7 to 1.0, where the positive value indicates that the moment is along the inflow direction (Fig. 5b). Clear periodicity is also observed at a frequency of $f_T = 0.22$ Hz in the zoomed-in plot. In comparison to the edgewise bending moment, the time series of flapwise bending moment displays more jitters (i.e., less smooth). This behavior may be associated with the discrepancy in the sensitivity of the edgewise and flapwise moments to the change of turbine structural configuration (i.e., the change of blade positions with respect to the turbine tower), which will be further supported through spectral analysis below.

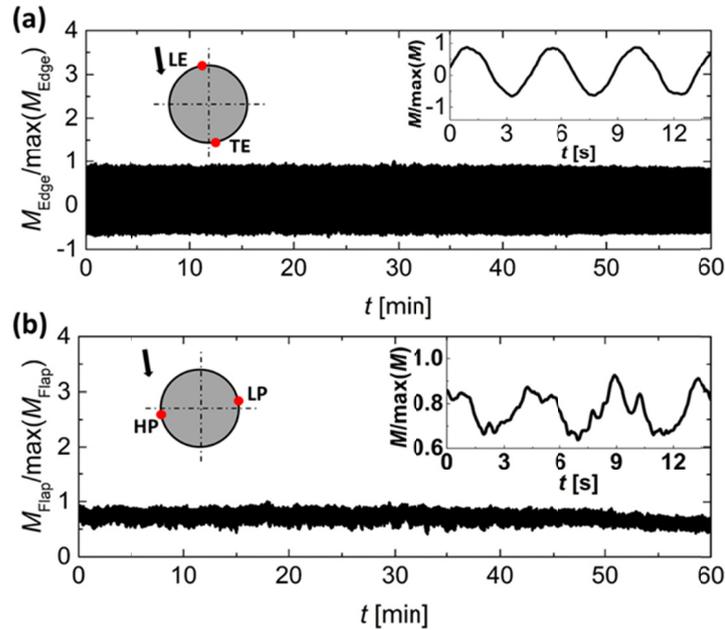

Figure 5: Time series of the normalized blade root bending moments of a rotating blade during the selected period. (a) Normalized edgewise blade root bending moment ($M_{Edge}$/max($M_{Edge}$)); (b) normalized flapwise blade root bending moment ($M_{Flap}$/max($M_{Flap}$)); The sketch in each subfigure shows the locations of the strain gauges installed at LE and TE regions, and over HP and LP surfaces of the blade root cross-section used to calculate the edgewise and flapwise blade root bending moments. The arrow indicates the direction of the resultant velocity at the blade

root cross-section. The zoomed-in plot in each subfigure shows the normalized blade root bending moments of a rotating blade over a short period to present more details.

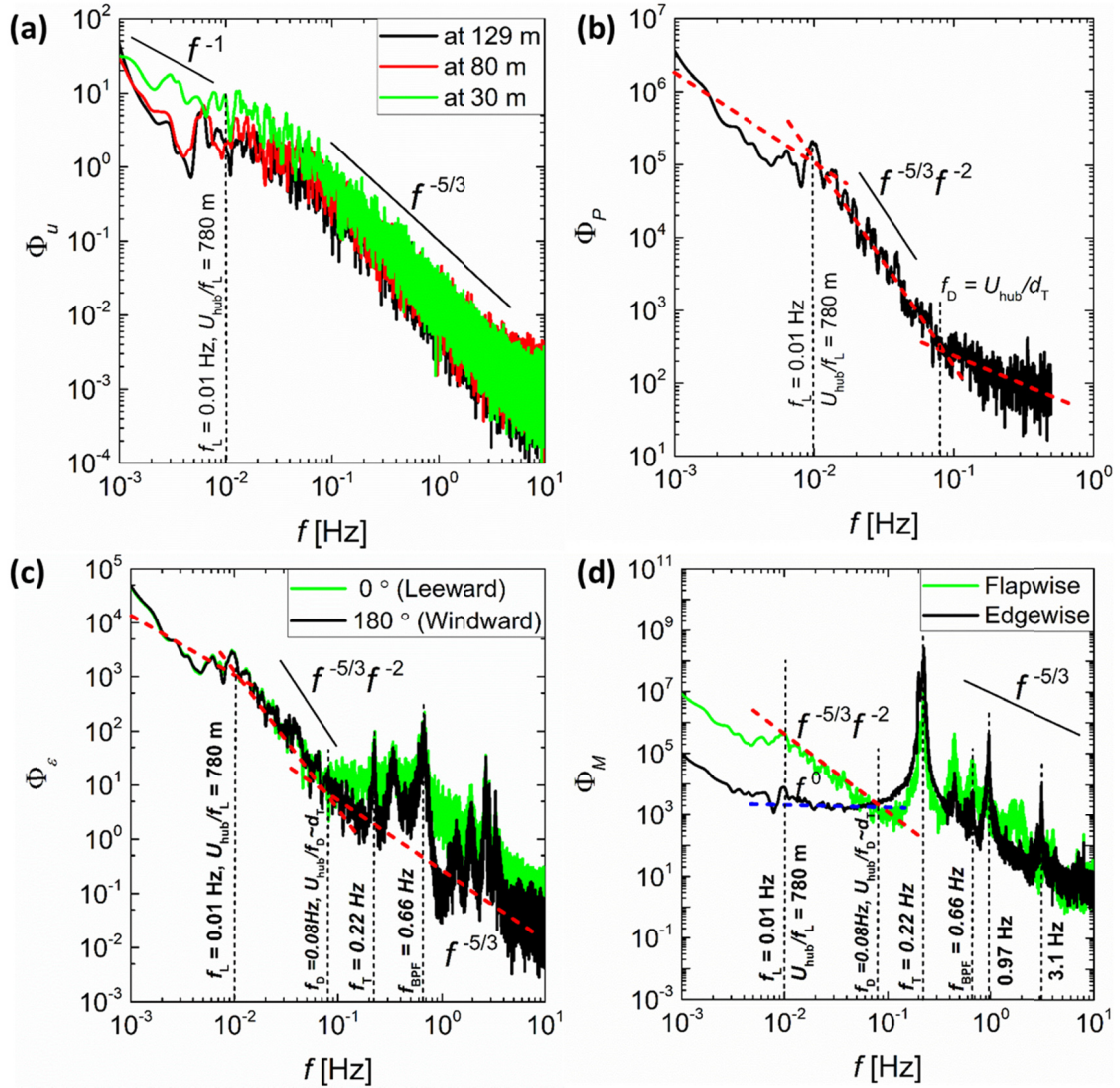

Figure 6: Power spectra of wind turbine quantities. (a) Streamwise components of the turbulent inflow at different elevations: 30 m (turbine bottom), 80 m (hub height), and 129 m (top tip). (b) wind turbine power; (c) foundation strains (0° and 180°, where 0 ° indicates direct north); (d) flapwise and edgewise blade root bending moments. The dashed lines present the least square fitting curves at lower, intermediate and higher ranges of frequencies to identify their slopes.

To obtain a better understanding of the blade structural response under turbulent inflow at different temporal scales, spectral analysis is conducted for the blade root flapwise and edgewise bending moments and other important wind turbine quantities, including turbine power and foundation strain, for comparison. Figure 6(a) shows the power spectra of the streamwise turbulent inflow velocity at three representative elevations: bottom tip (30 m), hub height (80 m), and top tip (129 m). Similar trends are observed across different elevations, all of which exhibit a clear $f^{-5/3}$ behavior in the inertial range of the spectra. Moreover, as shown in Fig. 6(b), (c), and (d), distinct dynamic responses are observed for the three frequency ranges, demarcated by two critical frequencies referred to as the boundary layer frequency ($f_L$) and the rotor frequency ($f_D$) hereafter. The former is about 0.01 Hz in the present study, corresponding to a length scale on the order of the boundary layer thickness $\delta$ [28], i.e., $U_{hub}/f_L = 780$ m $\sim O(\delta)$. The latter is associated with the time scale of flow across the rotor span, defined as the ratio of the hub-height velocity and the turbine rotor diameter ($d_T$), i.e., $f_D = U_{hub}/d_T$, which is about 0.08 Hz here. Correspondingly, the three frequency ranges are referred to as the lower (i.e., $f < f_L$), intermediate, ($f_L < f < f_D$) and higher ranges of frequencies (i.e., $f > f_D$) hereafter.

In the lower range of frequencies, blade structural response spectra are dictated by large-scale turbulent structures in the atmospheric boundary layer, which follows -1 power law in the fluctuating velocity spectra (Fig. 6a), according to [24,29]. Note that similar trends are also observed in the spectra of power production and foundation strain.

In the intermediate frequency range, as shown in Fig.6(d), a decay in the form of $\Phi_M = G(f)\Phi_U$, where $G(f) \propto (\sim)f^{-2}$ is the transfer function, is observed for the flapwise blade root bending moment. The decay may be associated with the damping effect of wind turbine structures. The same trends are also found in the spectra of power and foundation strain, as shown in the intermediate ranges in Fig. 6(b) and (c). Similar findings were reported by Chamorro et al. [24]. In comparison to the flapwise bending moment, the spectrum of the edgewise moment is much flatter (almost following $f^0$) in the intermediate range. This trend indicates that the edgewise bending moment is little influenced by the interaction of the turbulent inflow and the turbine within this scale range.

In the higher range of frequencies, the flapwise bending moment spectrum has the most dominant peak at the rotational frequency ($f_T = 0.22$ Hz), in contrast to those at $3f_T$ (i.e., $f_{BPF} = 0.66$ Hz) observed in the foundation strain spectra, as shown in Fig. 6(c) and in agreement with the conclusion derived from numerical simulations reported by Lee et al. [12]. This trend suggests the flapwise bending moment at the blade root is less affected than the foundation strain by the change of turbine structural configuration (i.e., the change of blade positions with respect to the turbine tower). Moreover, in comparison to the flapwise bending moment, the edgewise bending moment yields even lower energy at $2f_T$ and $3f_T$. This result suggests the edgewise moment is less sensitive to the change of individual blade position with respect to the tower, consistent with the observation of a smoother curve in Fig. 5(a) compared with that of flapwise moment in Fig. 5(b). Nevertheless, the edgewise quantity exhibits strong peaks at 0.97 Hz and 3.1 Hz. Such peaks are associated with the undamped natural frequency of the collective blade pitch control and its multiple [30]. This trend indicates that the edgewise bending moment is more prone to the influence of blade pitch control than the flapwise moment. It is also worth noting that no obvious peaks are observed in the spectrum of the power output in the high-frequency range, suggesting the turbine power is relatively insensitive to the change of turbine structural configurations.

**3.2 Blade structural response at different locations and azimuth angle positions**

Following the analysis of flapwise and edgewise bending moments at the blade root, the structural response of the blade at different circumferential locations (i.e., HP, LP, LE and TE regions) of a cross-section and different radial positions (i.e., blade root (RT), 25% $R$, 37.5% $R$ and 45% $R$, where $R$ represents the radius of the turbine rotor, 48 m) along the blade are investigated. Figure 7 compares the power spectra of blade strain measurements at different circumferential locations of the blade root cross-section. In the lower and intermediate frequency ranges, the HP spectrum matches well with the LP spectrum with a slope of $f^{-5/3}f^{-2}$, but differs from those of LE and TE. Specifically, LE and TE strain spectra show significantly lower slopes, with LE being almost flat, indicating that the LE and TE regions of blades are much less responsive to the turbulent inflow within these scale ranges. This trend is consistent with the flapwise bending moment being the dominant component of blade structural response, potentially due to the design of the blade cross-sectional profile resulting in higher rigidity in the edgewise direction. Additionally, the slight difference observed in the LE and TE strain spectra may be associated with the flow separation in the TE region, which reduces the correlation of the flow in that region with the external flow. In the higher frequency range, the spectra at all circumferential positions exhibit a similar slope of $f^{-5/3}$, and they share the dominant peak at $f_T = 0.22$ Hz and a second peak at $2f_T$. However, the LE/TE spectra exhibit higher peaks at 0.97 Hz and 3.1 Hz, consistent with the edgewise bending moment spectra, providing further support for the comparatively intense effect of blade pitching on blade deformation in the edgewise direction.

Figure 8 compares the power spectra of blade deformation at different radial positions. The spectra at all positions yield similar trends, for both HP and LP measurements. Note that no LE and TE measurements are available except at the blade root. In all frequency ranges, as shown in Fig. 8(a), the magnitude of the HP strain power spectrum increases in magnitude as the position extends outward from the blade root, likely due to the increased elasticity at the outer positions of the blades. In the intermediate range of frequencies, the slopes of the HP spectra decrease in magnitude at outer radial positions (i.e., the normalized slopes by the value at the blade root are 1.00, 0.95, 0.90, and 0.87 for RT, 25% $R$, 37.5% $R$, and 45% $R$, respectively), potentially also due to the increased blade elasticity at outer positions [18]. In the higher range of frequencies, the spectra of blade strain measured at different positions peak at the rotation frequency of $f_T$ = 0.22 Hz and its higher-order harmonics. In comparison with the power spectra of HP strains, the power of the LP strain is the highest for all frequencies at 37.5% $R$, instead of at the outermost location (i.e., 45% $R$ in the present study). This discrepancy may be caused by the difference of local flow separation along the blades, which can induce additional blade vibration [31,32].

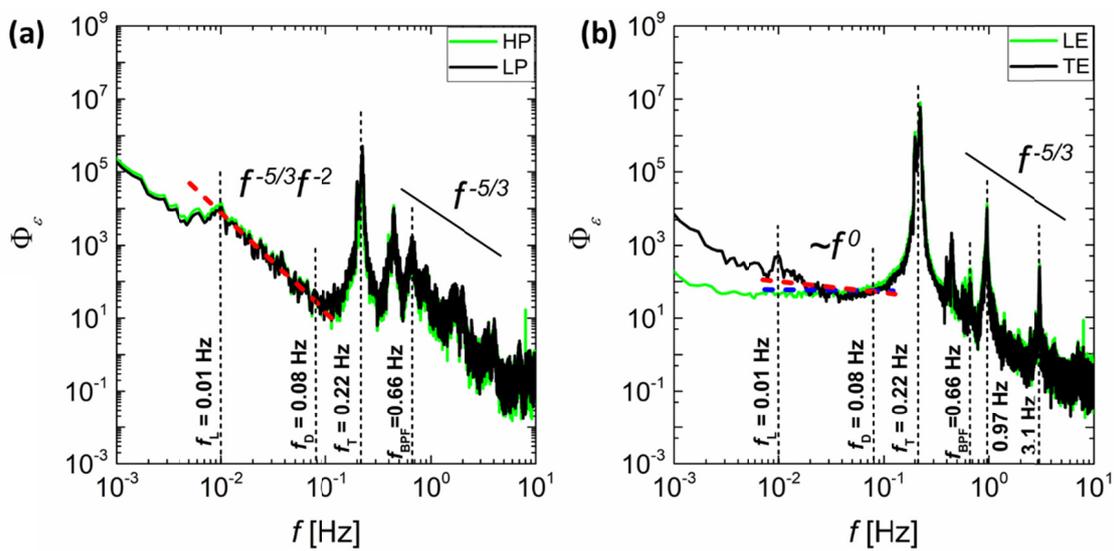

Figure 7: Comparison of power spectra of blade strain ($\Phi\varepsilon$) at different circumferential locations of the cross-section at the blade root (RT). (a) High-pressure (HP) and low-pressure (LP) surfaces; and (b) leading-edge (LE) and trailing-edge (TE) regions.

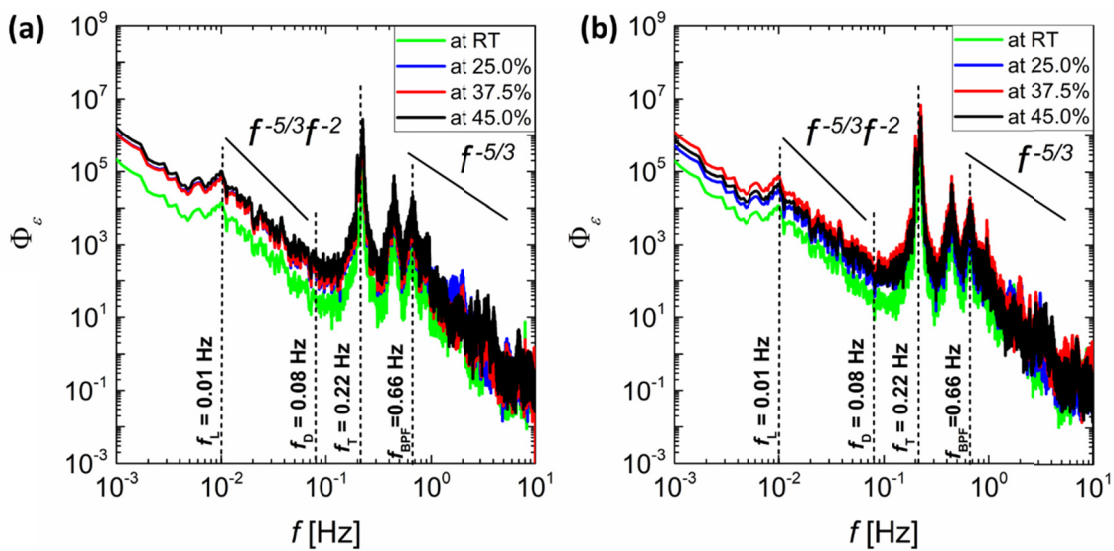

Figure 8: Comparison of the power spectra of blade strain ($\Phi\varepsilon$) at different radial locations. (a) High-pressure (HP) and (b) low-pressure (LP) surfaces.

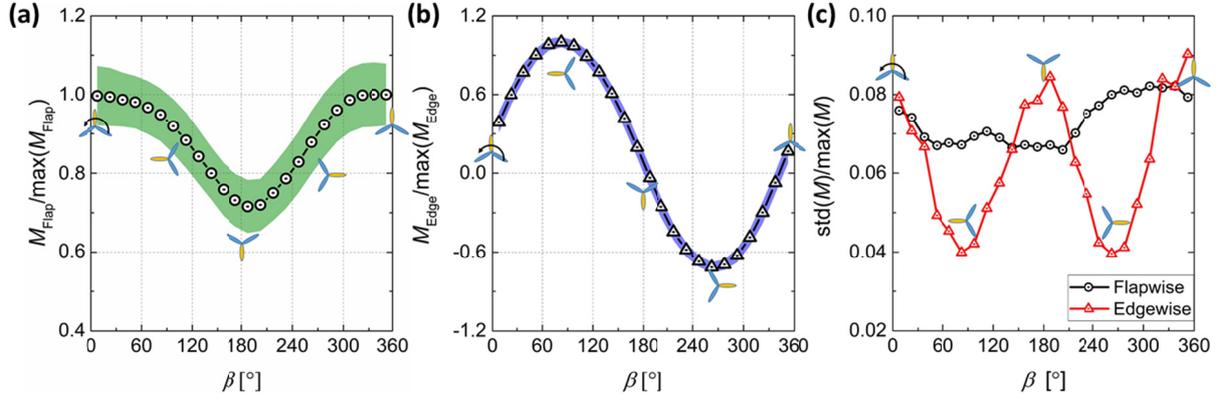

Figure 9: Effect of azimuth angle on blade root bending moments of a rotating blade. (a) Normalized flapwise blade root bending moment ($M_{Flap}/\max(M_{Flap})$); (b) normalized edgewise bending moment ($M_{Edge}/\max(M_{Edge})$); and (c) normalized fluctuation (i.e., std($M$)/max($M$)) of flapwise and edgewise bending moments. The symbols indicate the mean values of the normalized results in each azimuth angle interval (15 °) during the selected period while the color bars represent the fluctuations (i.e., one standard deviation). The blade highlighted in yellow indicates the blade azimuth position.

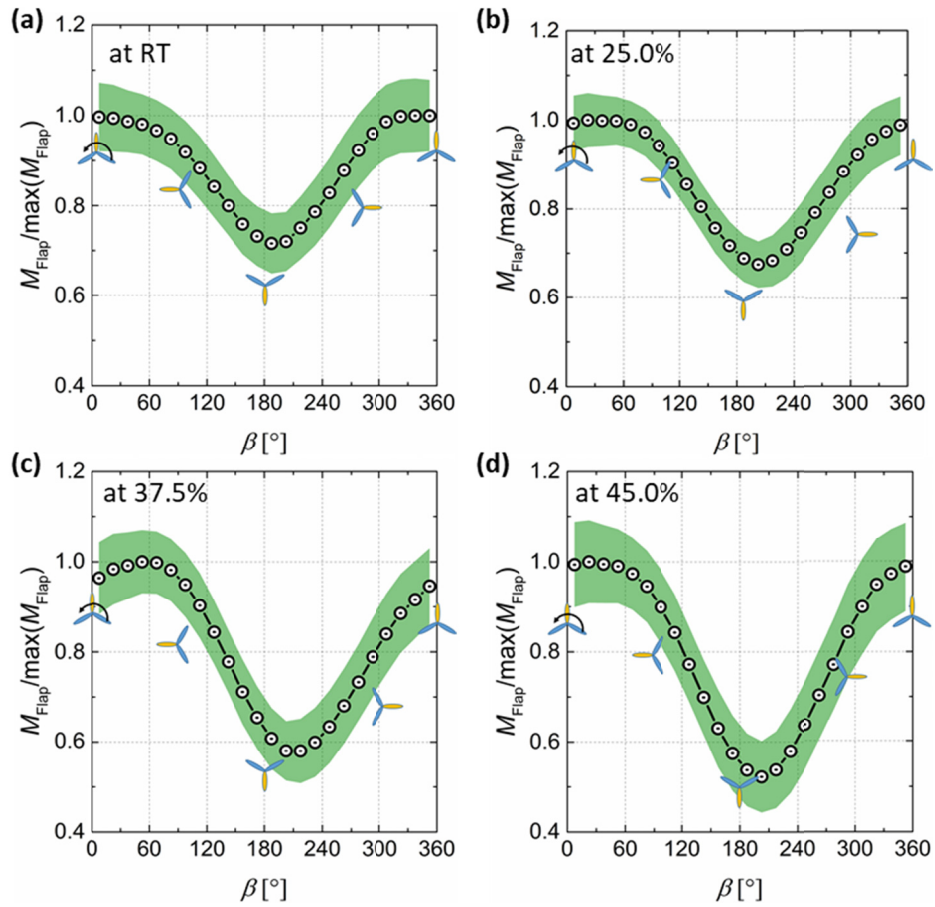

Figure 10: Comparison of flapwise blade root bending moment as a function of azimuth angle among different radial positions. (a) At blade root (RT); (b) at 25 % $R$; (c) at 37.5% $R$; and (d) at 45% $R$ of the turbine blade. The blade highlighted in yellow indicates the blade azimuth position.

The blade structural response at different azimuth angles is also characterized. Compared with the flapwise bending moment, the edgewise blade root bending moment has an overall much lower mean value (i.e., mean($M_{\text{Flap}}$)/mean($M_{\text{Edge}}$) = 9.7) and fluctuation (i.e., std($M_{\text{Flap}}$)/std($M_{\text{Edge}}$) = 1.4) for all azimuth angle positions. This trend suggests the turbulent inflow has a greater influence on the flapwise bending moment than that on the edgewise moment, consistent with the trend observed in the spectral analysis. As shown in Fig. 9(a), the mean flapwise blade root bending moment has maxima and minima when the blade points up and down, respectively. The same trends are observed in the flapwise fluctuations, as shown in Fig. 9(c). The variations in the mean value and fluctuation of the flapwise bending moment at different azimuth angles are mainly caused by the wind shear, i.e., the change in wind speed across different elevations. Note that the lowest blade bending moment is not precisely at 180 º due to the rotational effect. In comparison with the flapwise bending moment, a different trend is observed in the edgewise moment as shown in Fig. 9(b). Its mean reaches the largest magnitude when the blade is at horizontal positions and the minima when the blade points up or down, aligning with the direction of gravity, suggesting blade weight plays a dominant role in the variation of this quantity. Note that the positive and negative values of the edgewise moment indicate that the blade rotates along and against the direction of gravity, respectively. In contrast to the mean, the fluctuation of the edgewise moment, as displayed in Fig. 9(c), shows the opposite trend. Particularly, the highest fluctuations are observed when the blade points up or down, indicating an augmentation of blade vibration associated with wind shear or blade-tower interaction [33].

In addition, the effect of azimuth angle on flapwise bending moment is investigated at different radial locations, as summarized in Fig. 10. The general trend of flapwise moment variation with azimuth angle are similar across different locations, but the magnitude of variation, characterized by the difference between the maximal and minimal $M_{\text{Flap}}$, increases with outer radial positions. This trend is likely caused by the increasing discrepancy of wind velocity across different azimuth angles at outer positions due to wind shear. The fluctuation of $M_{\text{Flap}}$ reaches the highest for all azimuth angles in the outmost radial position, consistent with the increase of blade elasticity with radial position as suggested earlier. Remarkably, the minimal $M_{\text{Flap}}$ seems to take place at an azimuth angle farther away from 180 º compared to that at blade root potentially due to an enhanced rotational effect at outer positions.

### 3.4 Temporal correlation between the turbulent inflow and blade structural response

To illustrate the linkage between the turbulent inflow and the blade response, delay-dependent Pearson correlation is employed, i.e., $R_{A,B}(\tau) = \{cov[A(t),B(t-\tau)]\}/\sigma_A\sigma_B$, where $A$ and $B$ represent the two parameters, $cov$ represents covariance, and $\tau$ is the time lag, to calculate the temporal correlation. Figure 11(a) shows the delay-dependent correlations of the streamwise velocity component of the turbulent inflow with the flapwise ($R_{U,M\text{Flap}}$), and edgewise ($R_{U,M\text{Edge}}$) bending moments at the blade root. The value of $R_{U,M\text{Flap}}$ exhibits a distinct peak at a time delay of ~ 28 s, matching approximately the advection time of flow structures from the met-tower (i.e., where the turbulent inflow is measured) to the turbine rotor plane [3]. In contrast, $R_{U,M\text{Edge}}$ does not yield any appreciable peaks and remains near zero regardless of the time delay, suggesting little correlation between the turbulent inflow and the edgewise moment at the blade root. As shown in Fig. 11(b), similar trends for $R_{U,M\text{Flap}}$ are observed at other radial positions. However, the $R_{U,M\text{Flap}}$ at the blade root yields significantly higher values than those at other positions, which show little difference among them. This observation can be explained by the fact that the turbulent inflow probed at the hub (i.e., the $U$ used to correlate with the blade response) is better correlated with the flow interacting with the blade root around the hub than those at outer radial positions. Additionally, the correlation between $M_{\text{Flap}}$ and $M_{\text{Edge}}$ at the blade root exhibits strong fluctuations from -0.7 to 0.7 with changing time delay (Fig. 11c), which is associated with the change of the sign of $M_{\text{Edge}}$ with changing blade position, as shown in Fig. 9(b).

To further clarify the trends in $R_{U,M\text{Flap}}$ and $R_{U,M\text{Edge}}$, these correlations are evaluated with low-pass filtered quantities, as shown in Figs. 11(d), (e) and (f). Specifically, the Chebyshev Type I filter is used to suppress the signals with a frequency above the rotor frequency ($f_{\text{D}}$), and the corresponding filtered quantities are denoted as $\widetilde{U}^{>f_D}$, $\widetilde{M}_{\text{Flap}}^{>f_D}$ and $\widetilde{M}_{\text{Edge}}^{>f_D}$. In general, the correlations based on these low-pass filtered quantities show similar trends but with improved correlation values. Particularly, the $R_{\widetilde{U},\widetilde{M}\text{Flap}}$ shows an increase of maximum correlation to 0.70 (Fig. 11d). Remarkably, low-pass filtering also substantially reduces the discrepancy of $R_{\widetilde{U},\widetilde{M}\text{Flap}}$ across different radial locations (Fig. 11e), suggesting the influence of radial location on $R_{\widetilde{U},\widetilde{M}\text{Flap}}$ resides mainly in the higher range of frequencies. Moreover, after the low-pass filtering, the correlation of flapwise and edgewise bending moments ($R_{\widetilde{M}\text{Edge},\widetilde{M}\text{Flap}}$) is significantly enhanced and decays monotonically and slowly with increasing time delay (Fig. 11f), which is a typical manifestation of mechanical coupling between $\widetilde{M}_{\text{Flap}}$ and $\widetilde{M}_{\text{Edge}}$.

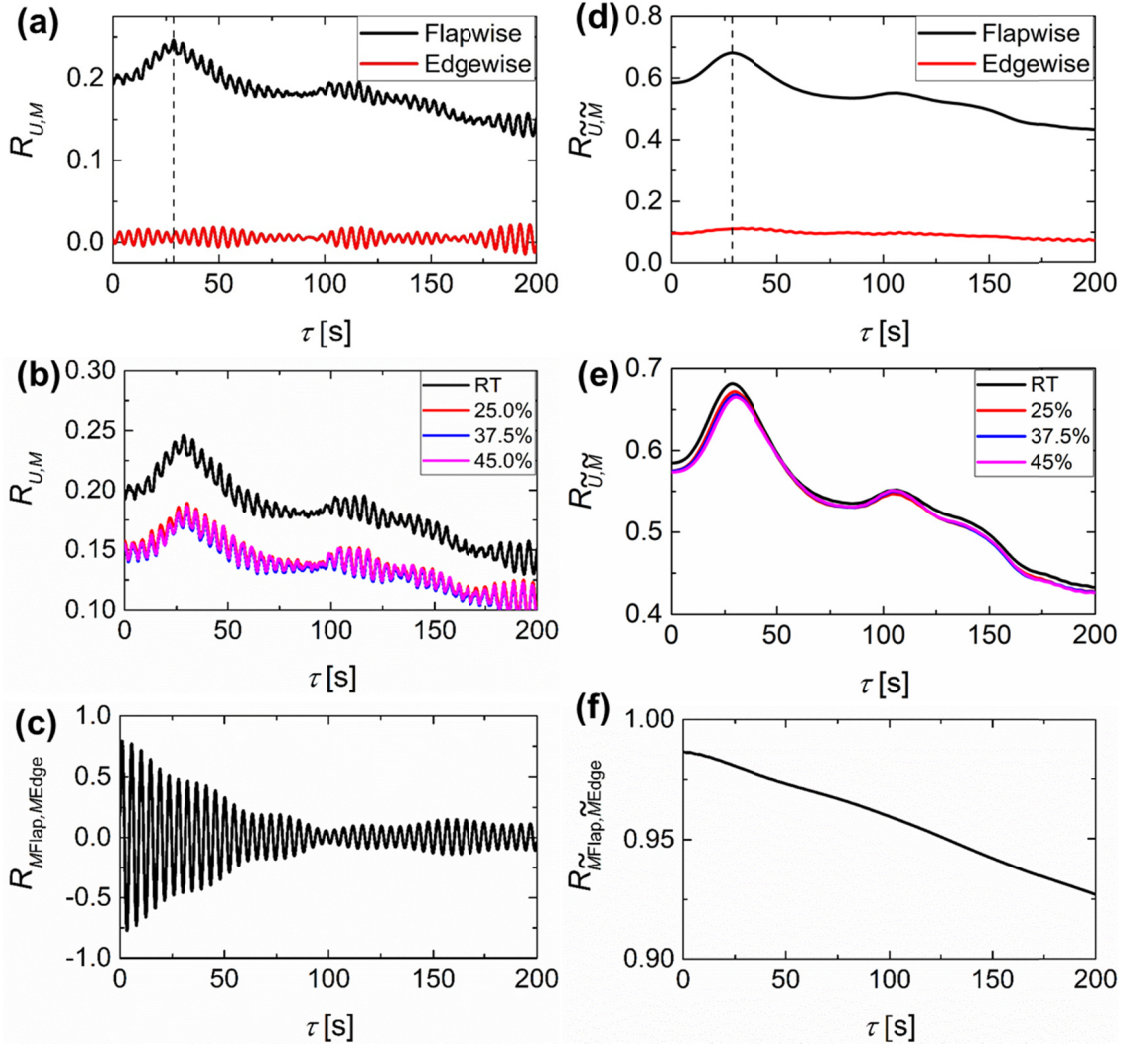

Figure 11: Delay-dependent correlations between turbulent inflow and blade structural response. (a) Correlation of the turbulent inflow streamwise velocity component and flapwise/edgewise root bending moment; (b) correlation of the turbulent inflow streamwise velocity component and flapwise bending moment at different radial locations; (c) correlation of flapwise and edgewise root bending moments; and (d), (e) and (f) are correlations <of the quantities shown in (a), (b) and (c), respectively, with the frequencies above $f_D = 0.08$ Hz filtered out.

To further characterize the influence of different scales, temporal correlations between turbulent inflow and blade structural response are evaluated using band-pass filtered quantities. Here we divide the scale range into five different bands, i.e., $<f_L$, $[f_L, f_D]$, $[f_D, f_T]$, $[f_T, f_{BPF}]$, and $>f_{BPF}$. The first two correspond to the lower frequency and the intermediate frequency range introduced in the spectral analysis. To highlight the influence of blade rotation, the higher frequency range in the spectral analysis is further divided into three bands using the rotational frequency (i.e., $f_T$) and the blade passing frequency (i.e., $f_{BPF} = 3f_T$). The correlation peaks (according to different time delays) evaluated using the band-pass filtered quantities (i.e., $\max(R_{\widetilde{U},\widetilde{M}\text{Flap}})$, $\max(R_{\widetilde{U},\widetilde{M}\text{Edge}})$, and $\max(R_{\widetilde{M}\text{Edge},\widetilde{M}\text{Flap}})$) are summarized in Fig. 12. The filtered turbulent inflow and flapwise bending moment at the blade root exhibit a strong correlation in the scale range below the boundary layer frequency ($f_L$) with a correlation coefficient above 0.9. Such correlation decays slightly in the range of [$f_L, f_D$], and drops rapidly to around 0.3 when the scale rises above $f_D$. Similar trends are observed for the correlations between the filtered turbulent inflow and the edgewise bending moments. However, in comparison to $\max(R_{\widetilde{U},\widetilde{M}\text{Flap}})$, $\max(R_{\widetilde{U},\widetilde{M}\text{Edge}})$ yields lower values, particularly in the scale range above $f_D$. Specifically, $\max(R_{\widetilde{U},\widetilde{M}\text{Edge}})$ drops to almost zero while $\max(R_{\widetilde{U},\widetilde{M}\text{Flap}})$ has values around 0.3. These trends are generally consistent with those from our spectral analysis in Section 3.1, suggesting that the blade responds strongly to turbulent motions below the rotor frequency and for turbulence at higher frequencies, the response occurs primarily in the flapwise direction. Interestingly, the variation of the correlation between the flapwise and edgewise bending moments at the blade root is not monotonic with respect to the scale change. Specifically, $\max(R_{\widetilde{M}\text{Edge},\widetilde{M}\text{Flap}})$ exhibits a minimum in the range of [$f_D, f_T$], and shows a high correlation (i.e. >0.7) at both lower and higher ranges. This nonmonotonic behavior can be explained by the presence of two distinct mechanisms that influence the correlation between flapwise and edgewise bending moments at different ranges of scales. In the range below $f_D$, the blade structural response is dominated by large-scale turbulent motions which force similar flapwise and edgewise behaviors. In contrast, in the range above $f_T$, the correlation between flapwise and edgewise moments is primarily dictated by the structural deformation caused by turbine rotation. Consequently, in the range between $f_D$ and $f_T$, the mechanisms compete with each other, resulting in the drop of $\max(R_{\widetilde{M}\text{Edge},\widetilde{M}\text{Flap}})$.

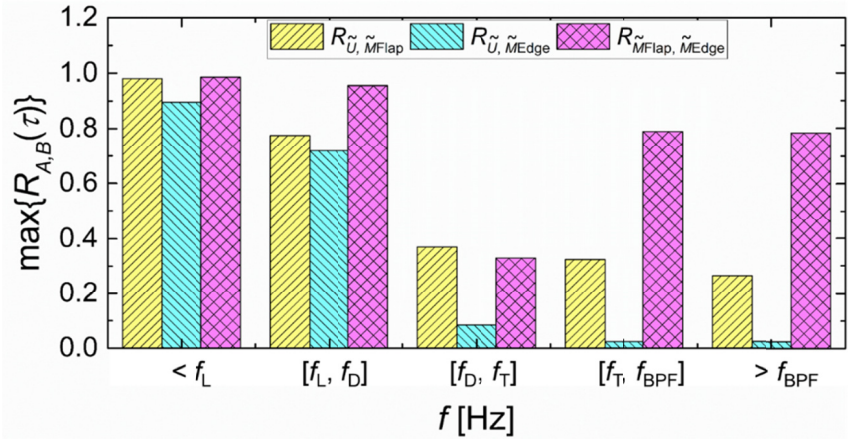

Figure 12: Scale-to-scale correlation among the streamwise velocity component of the turbulent inflow, and the flapwise and edgewise blade bending moments at the blade root.

## 4. Conclusions and Discussion

The present study systematically investigates the influence of turbulent inflow on the blade structural response of a utility-scale wind turbine using our unique facility available at the Eolos Wind Energy Research Station. A representative one-hour dataset under a stable atmosphere is selected for the characterization, including the turbulent inflow data measured from the meteorological tower, high-resolution blade strain measurement at different circumferential and radiation positions along the blade, and the wind turbine operational conditions. The main findings are summarized as follows.

Firstly, the spectral analysis of blade bending moment and supplementary comparison with turbulent inflow velocity, turbine power and foundation strain reveals three different scale ranges that lead to distinct blade response behaviors, corresponding to the modulations due to large-scale eddies in the atmosphere (i.e., lower frequency range), time scales associated with the flow across the rotor span (i.e., higher frequency range), and those in between (i.e., intermediate frequency range). A strong energy peak at rotation frequency ($f_T$) appears in the blade root bending moment spectrum, in contrast to the spectrum of foundation strain which peaks at blade passing frequency ($f_{BPF} = 3f_T$). Moreover, in comparison to the flapwise bending moment, the edgewise quantity exhibits strong peaks associated with the undamped natural frequency of the collective blade pitch control and its higher-order harmonics. Secondly, a comparison of blade structural response at different circumferential and radial locations shows strong variability which is attributed to the variation of blade elasticity and the local flow condition/separation across different blade regions. The bending moment of the blade also varies substantially according to its azimuth position. Specifically, the flapwise moment reaches a maximum as the blade points straight up (i.e., 0º position) due to the effect of wind shear, while its minimum occurs at positions deviating from the blade straight down position (i.e., 180º position), with increasing deviation at outer radial positions of the blade, potentially due to blade rotation. In contrast, the variation of the edgewise moment with the azimuth angle is primarily dominated by the blade weight, i.e., the magnitude peaks when the blade is horizontal. Additionally, the turbulent inflow exhibits a stronger influence on the flapwise moment than the edgewise, manifested by significantly higher mean and fluctuation of the flapwise moment. Thirdly, the temporal correlation of the turbulent inflow with the blade response at the root displays a strong correlation for the flapwise moment at a time delay matching the advection time of the large-scale flow structures, while little correlation is observed for the edgewise response. Moreover, the temporal correlation at different scale (frequency) ranges shows both flapwise and edgewise moments correlate strongly with the inflow signal below the rotor frequency (i.e., $f_D$), and their correlations decay rapidly with increasing frequency due to the increased effect of blade rotation and other high-frequency turbine dynamic characteristics in the higher range. Interestingly, the blade shows dominant response to turbulence flapwise than edgewise particularly in high frequencies.

Our work provides first-hand information about the wind turbine blade structural behavior under the turbulent inflow utility-scale wind turbines experience during real-world operation. Such information can be used for validating the corresponding numerical results [12,13,21] and experimental work [8]. Furthermore, the data presented in the present study can be implemented to evaluate and enhance the aeroelastic models/codes, such as FAST [12] and nonlinear-based WindGRAR [18], for wind turbine applications. Additionally, the derived findings regarding the blade structure behaviors under turbulent inflow allow the development of new control strategies and blade design for large-scale wind turbines (i.e., > 2.0 MW) to mitigate the structural impact and thereby extend turbine life as the unit continues upsizing. Specifically, numerical simulation for blade design requires a high temporal resolution above the blade passing frequency, $f_{BPF}$, to consider the modulations induced by the turbulent flow across the blade sections, while a comparatively lower temporal resolution on the order of rotor frequency (i.e., $f_D$) is adequate for a survey with turbine power fluctuation as primary interest. The high and low pressure (suction) sides of blades exhibit significantly stronger turbulence-induced deformation compared with the leading- and trailing-edge regions. Therefore, the future large-scale turbine blade should consider effective design innovations to mitigate such disparity of blade response at different circumferential locations. Our study suggests that the deformation of blade is not only influenced by its elasticity, but also affected by the local flow separation on its suction side surface. Accordingly, flow control devices (e.g., vortex generators) installed on the suction side to reduce flow separation can also mitigate the associated structural deformation. Our study provides detailed information of the blade response under practical turbine operational conditions. Such conditions incorporating the effect of wind shear, turbulence and blade rotation, are not usually considered in the standard analysis of blade loading conducted by the blade manufacturer. We expect this detailed information, particularly regarding the variability of blade response at different radial locations under different azimut angles, can help the blade manufacturer provide more accurate estimates of blade lifespan.

It should also be cautioned that our present study, though generally applicable to utility-scale variable-speed turbines, is conducted under stable stratification conditions. Variations in atmospheric stability could substantially change the wind profile and turbulence characteristics that may influence our findings on the blade response to turbulence. Future studies will look into this subject under neutral and unstable atmospheric stratifications using our database.

## Conflict of Interest

The authors declare that there is no conflict of interest regarding the publication of this paper.

## Acknowledgments

This work was supported by the National Science Foundation CAREER award (NSF-CBET-1454259), Xcel Energy through the Renewable Development Fund (grant RD4-13) as well as IonE of University of Minnesota. The authors would thank Christopher Milliren, Associate Engineer at St. Anthony Falls Laboratory, for the fruitful discussion regarding the Eolos database.